\documentclass{article}
\usepackage{spconf,amsmath,graphicx}
\usepackage{booktabs}
\usepackage{graphics}
\usepackage{enumitem}
\usepackage{xcolor}

\title{Towards Lightweight Applications: Asymmetric Enroll-Verify Structure for Speaker Verification}
\name{Qingjian Lin$^\star$, Lin Yang$^\star$, Xuyang Wang$^\star$, Xiaoyi Qin$^{\dag}$, Junjie Wang$^\star$, Ming Li$^\dag$}
\address{$^\star$AI Lab, Lenovo Research, Beijing, China \\
         $^\dag$Data Science Research Center, Duke Kunshan University, Kunshan, China \\
}
\begin{document}
%\ninept
%
\maketitle
% server, device
\begin{abstract}
With the development of deep learning, automatic speaker verification has made considerable progress over the past few years. However, to design a lightweight and robust system with limited computational resources is still a challenging problem. Traditionally, a speaker verification system is symmetrical, indicating that the same embedding extraction model is applied for both enrollment and verification in inference. In this paper, we come up with an innovative asymmetric structure, which takes the large-scale ECAPA-TDNN model for enrollment and the small-scale ECAPA-TDNNLite model for verification. As a symmetrical system, our proposed ECAPA-TDNNLite model achieves an EER of 3.07\% on the Voxceleb1 original test set with only 11.6M FLOPS. Moreover, the asymmetric structure further reduces the EER to 2.31\%, without increasing any computational costs during verification.

\end{abstract}
\begin{keywords}
lightweight speaker verification, asymmetric enroll-verify structure, ECAPA-TDNNLite
\end{keywords}
\section{Introduction}
\label{sec:intro}
Automatic speaker verification (ASV) refers to the process of verifying a user's identity based on the voiceprint~\cite{1454425,poddar2018speaker}. Classification-based ASV systems generally consist of two stages: in the enrollment stage, ASV systems extract a fixed-dimensional speaker embedding according to the user's voice; then in the verification stage, given the unknown speech, speaker embeddings are extracted and compared with the enrolled one. A preset threshold makes the final decision to accept or reject the speech.

In the past years, performance of ASV has made significant improvement due to the successful application of deep neural networks (DNN)~\cite{8461375,luo18b_interspeech,garciaromero20_odyssey,thienpondt2021integrating}. However, the computational complexity also increases accordingly. For devices like mobile phones and IoT terminals, it matters to develop low-latency models with limited resources, and the task has attracted much attention. For example, \cite{cai18_odyssey} and~\cite{chung20b_interspeech} halve the number of channels and prepose strides to reduce computational requirements of ResNet34. \cite{koluguri2020speakernet} and~\cite{9413564} develop the lightweight models based on separable convolutions~\cite{Chollet_2017_CVPR}. \cite{zhu21_interspeech} applies binary neural networks to the task, while~\cite{8683443} utilizes knowledge distillation to guide the student model with the teacher model. They are all symmetrical systems.

In this paper, we propose an asymmetric structure at the system level, where models of different scales are separately employed in the enrollment and verification stages. Specifically, a large-scale model with higher accuracy and larger computational consumption is applied for enrollment, while a small-scale model balancing performance and inference latency executes during verification. As a result, the asymmetric structure achieves better performance than the small-scale model. We argue that it benefits lightweight applications due to the following reasons:
\begin{itemize}
  \setlength{\itemsep}{2pt}
  \setlength{\parsep}{0pt}
  \setlength{\parskip}{0pt}
  \item The DNN based embedding extraction model executes in the verification stage for most of the time, since users usually enroll their voices only once. Employing a large-scale model during enrollment does not significantly increase the overall computational complexity. 
  \item Users are less sensitive to the latency of enrollment. Besides, in some IoT application scenarios, feedbacks like ``enrollment success" can be presented immediately even though the device is still processing the enrolled speech, which improves the user experience.
  \item The asymmetric structure exactly meets the scenarios where speakers enroll their voices on the server while verifying identities on devices. The server generally owns more abundant resources and supports larger models. Symmetrical systems, however, limit the server side to small-scale models only, leading to poorer performance. % at the expense of sacrificing the total performance.
\end{itemize}
Another highlight of our paper is ECAPA-TDNNLite, a small-scale model based on ECAPA-TDNN~\cite{9414600,desplanques20_interspeech}. ECAPA-TDNNLite reduces computational costs by squeezing feature mapping sizes during calculation and employing separable convolutions instead, which reaches the balance between performance and inference latency.

% ECAPA-TDNNLite achieves an equal error rate (EER) of 3.07\% independently, with only 11.6M floating-point operations per second (FLOPS). Besides, our proposed asymmetric structure further reduces the EER by 25\% relatively, with no additional computational costs during verification. To our knowledge, the result is very competitive among recent lightweight systems.

The rest of this paper is organized as follows. The next section introduces details of the asymmetric structure. Section 3 describes the experimental setup and evaluation protocol. Results and discussions are presented in Section 4, while conclusions are drawn in Section 5.

\section{Asymmetric Structure}
\subsection{Overview}
\begin{figure}[t]
  \centering
  \includegraphics[width=\linewidth]{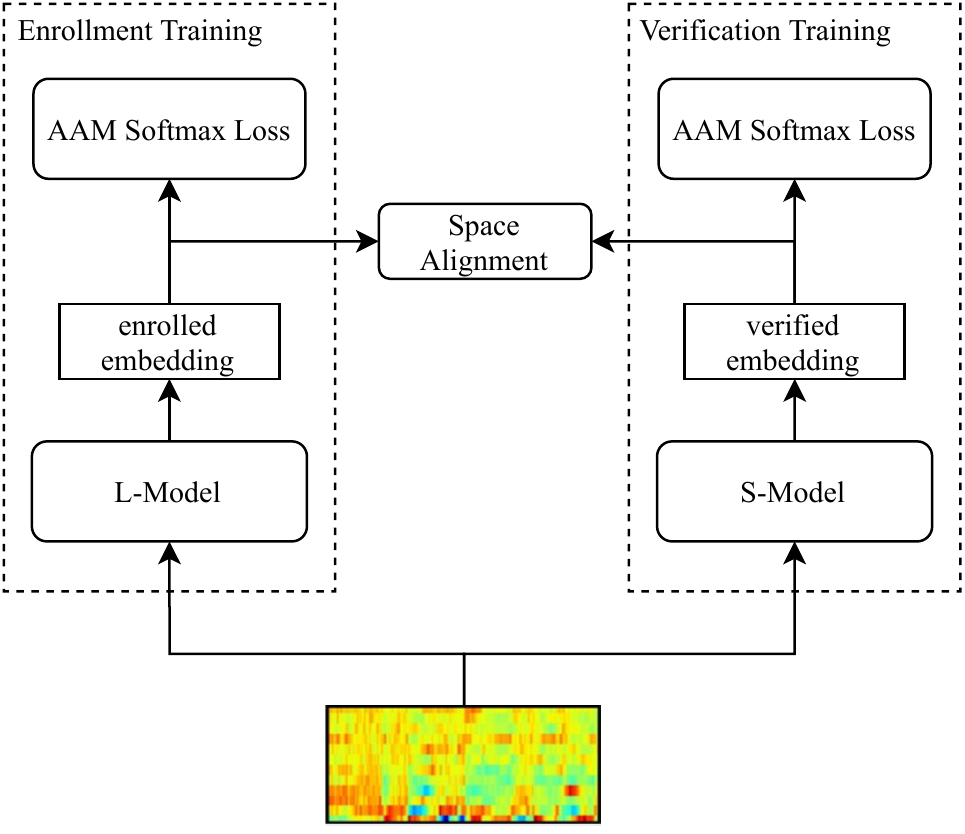}
  \vspace{-6mm}
  \caption{The training process of the asymmetric structure. Frame-wise input features are fed into the large-scale model and the small-scale model, respectively. AAM softmax loss is applied on both enrollment and verification sides, and an additional loss function aligns the speaker subspace bewteen the enrolled embeddings and the verified embeddings.}
  \label{fig:asymmetric}
\end{figure}

\label{sec:asymmetric}
The training process of the asymmetric structure is shown in Fig.~\ref{fig:asymmetric}. Input features like MFCCs are fed into the large-scale model (L-Model) and the small-scale model (S-Model), respectively. Then classification-based loss functions compute the loss between speaker embeddings and the corresponding ground-truth labels. Following popular configurations in ASV, we employ the additive angular margin softmax (AAM Softmax) loss~\cite{Deng_2019_CVPR}. Besides, an additional loss function is proposed to align the enrolled embedding and the verified embedding from the same input utterance.

In inference, the L-Model computes the enrolled embedding according to the target speaker's voice while the S-Model extracts the verified embedding given the unknown speech. It is worth noting that if the S-Model is used for both enrollment and verification, the whole system degrades to the embedding-level knowledge distillation solution like~\cite{8683443}.

% A simple cosine function measures similarity between the two embeddings. 

% In inference, we employ the L-Model to extract the enrolled embedding from the target speaker's voice. Then given unknown speech, the verified embeddings are extracted with the S-Model and compared with the enrolled one to generate similarity scores. A preset threshold finally converts the float scores to binary decisions.

\subsection{Space Alignment}
A key problem in the asymmetric structure is that the enrolled embedding and the verified one may be derived from different speaker subspaces, leading to mismatch in inference. Therefore, it is necessary to maximize similarity of the two embeddings in the training process, namely space alignment. It is similar to metric learning methods in ASV, and thus we investigate the angular prototypical (AP) loss in~\cite{chung20b_interspeech, heo2020clova} to our system with slight modification. Assume that a mini-batch contains $B$ samples. $\boldsymbol{e}_1, \boldsymbol{e}_2, ..., \boldsymbol{e}_B$ are enrolled embeddings extracted from the mini-batch, and $\boldsymbol{v}_1, \boldsymbol{v}_2, ..., \boldsymbol{v}_B$ are the verfied embeddings. The cosine similarity between $\boldsymbol{e}_i$ and $\boldsymbol{v}_j$ is
\begin{equation}
  \cos\theta_{i,j} = \frac{\langle\boldsymbol{e}_i, \boldsymbol{v}_j\rangle}{\|\boldsymbol{e}_i\|\|\boldsymbol{v}_j\|},
\end{equation}
and the loss function is defined as
\begin{equation}
  L_\text{AP} = -\frac{1}{B}\sum_{i=1}^B\log\frac{e^{w\cos\theta_{i,i} + b}}{\sum_{j=1}^Be^{w\cos\theta_{i,j} + b}}.
\end{equation}
In the original settings, $w$ and $b$ are trainable parameters. Here we view $w$ as a hyperparameter, and remove bias $b$ since it should have been canceled out by fraction. The AP loss aims at maximizing cosine similarity between $\boldsymbol{e}_i$ and $\boldsymbol{v}_i$ ($i=1, 2, ..., N$) while minimizing the similarity of the rest pairs. Therefore, utterances to form a mini-batch must be from different speakers. Let $L_\text{L-AAM}$ and $L_\text{S-AAM}$ be the two AAM loss functions, and the overall loss is formulated as
\begin{equation}
  L = L_\text{S-AAM} + L_\text{L-AAM} + \lambda \cdot L_\text{AP},
\end{equation}
where $\lambda$ is a scale factor to balance the losses.

\subsection{L-Model}
We employ ECAPA-TDNN as the L-Model, which is a new variant of the TDNN structure and has achieved promising success in ASV. Input MFCC features are fed into a Conv1D layer with stride $s=1$. The followings are three stacked SE-Res2Blocks. Each block contains a preceding dense layer, the dilated convolutions, a succeding dense layer and a squeeze-and-expansion (SE) layer~\cite{Hu_2018_CVPR}. The whole block is covered by a skip connection. Outputs from the three SE-Res2Blocks are concatenated over channel dimension and fed into another dense layer with 1536 units. Then the attentive statistics pooling (ASP) layer calculates weighted statistics over the temporal dimension, converting frame-wise feature mappings to utterance-wise vectors. The last dense layer reduces the vector dimension from 3072 to 192, generating output speaker embeddings. More details are reported in~\cite{desplanques20_interspeech}.

\subsection{S-Model}
\begin{figure}[t]
  \centering
  \includegraphics[width=\linewidth]{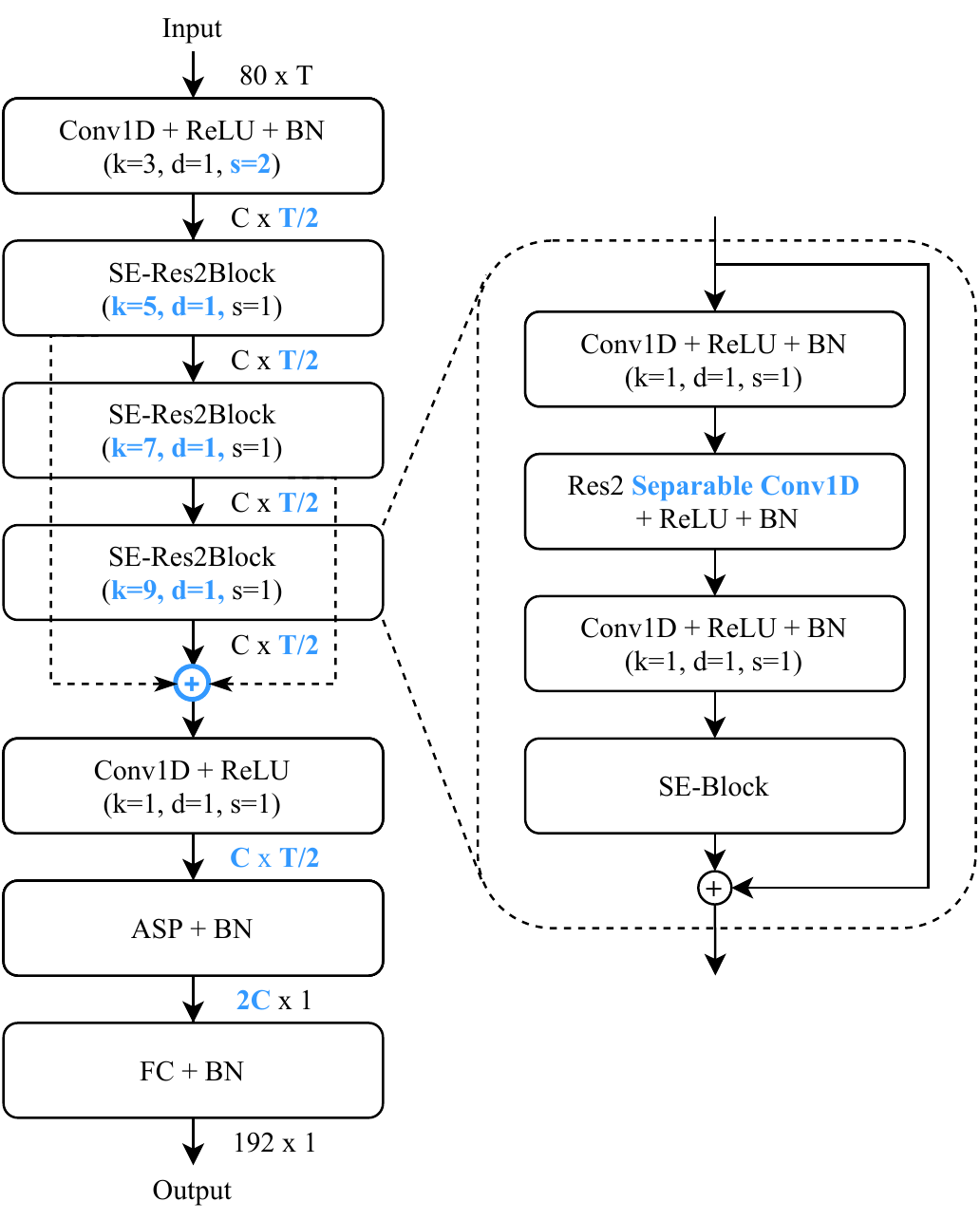}
  \vspace{-6mm}
  \caption{Network topology of ECAPA-TDNNLite. $k$ denotes kernel size, $d$ denotes dilation rate and $s$ denotes stride in Conv1D layers and SE-Res2Blocks. $C$ and $T$ are the channel and time dimensions of feature mappings, respectively. Modifications in comparison with ECAPA-TDNN are plotted in bold blue.}
  \label{fig:ECAPA_TDNNLite}
\end{figure}

Considering the requirements of real-life applications, the S-Model is expected to run under critical resource-limited conditions with only around 10M floating-point operations per second (FLOPS). Therefore, we extend the ECAPA-TDNN network to ECAPA-TDNNLite with the following modifications:
\begin{itemize}
  \setlength{\itemsep}{2pt}
  \setlength{\parsep}{0pt}
  \setlength{\parskip}{0pt}
  \item Change stride of the first Conv1D layer from 1 to 2. The configuration reduces the sequence length by half and thus cuts down 50\% computation, with slight degradation in performance.
  \item Replace dilated convolutions with separable convolutions in SE-Res2Blocks to further reduce the number of parameters, meanwhile maintaining the same receptive field. 
  \item Sum outputs of the three SE-Res2Blocks instead of concatenation. The concatenation operator results in high-dimensional feature mappings and relatively expensive computational costs for the following layers.
\end{itemize}
The whole network topology of ECAPA-TDNNLite is shown in Fig.~\ref{fig:ECAPA_TDNNLite}.

\begin{table}[b]
  \vspace{-5mm}
  \caption{Comparison between ECAPA-TDNNLite and recent works on EERs of Voxceleb1-O, the number of parameters, FLOPS and the real-time factor (RTF). The RTF is measured on a single core of Intel(R) Xeon(R) CPU E5-2650 v4 @ 2.20GHz with one thousand 10-second utterances.}
  \vspace{2mm}
  \resizebox{\columnwidth}{!}{%
  \begin{tabular}{@{\ \ }l@{\ }c@{\ \ }c@{\ \ }c@{\ \ }c@{\ \ }}
  \toprule
  \textbf{Model}           & \textbf{EER(\%)} & \textbf{\#Params} & \textbf{FLOPS} & \textbf{RTF} \\
  \midrule
  SpeakerNet-M~\cite{koluguri2020speakernet}     & 2.21             & 5.0M              & -              & -            \\
  VGG-M-40~\cite{chung20_odyssey}         & 4.64             & 4.0M              & 0.53G          & -            \\
  Thin ResNet-34~\cite{cai18_odyssey}   & 2.36             & 1.4M              & 0.93G          & -            \\
  Fast ResNet-34~\cite{chung20b_interspeech}   & 2.37             & 1.4M              & 0.45G          & -            \\
  Julien at al.~\cite{9413564}    & 3.31             & 238K              & 11.5M          & -            \\
  \midrule\midrule
  Julien at al.~\cite{9413564} (our impl.) & 3.32    & \textbf{243K}     & \textbf{11.6M} & $\text{7.0}{\times}\text{10}^\text{-3}$  \\
  ECAPA-TDNNLite           & \textbf{3.07}    & 318K              & \textbf{11.6M} & $\textbf{1.8}{\times}\textbf{10}^\textbf{-3}$     \\
  \bottomrule
  \end{tabular}
  }
  \label{tab:ecapa_tdnnlite}
\end{table}

\begin{table*}[t]
  \caption{Performance of the symmetrical and asymmetric structures on the Voxceleb1 dataset. Superscript $\dag$ means that the model is trained individually, while $\star$ indicates the model is trained in the asymmetric structure.}
  \vspace{2mm}
  \begin{tabular}{cllcccccc}
  \toprule
   & \multicolumn{2}{c}{\textbf{Model}}  & \multicolumn{2}{c}{\textbf{Voxceleb1-O}} & \multicolumn{2}{c}{\textbf{Voxceleb1-E}} & \multicolumn{2}{c}{\textbf{Voxceleb1-H}}\\
  \cmidrule(lr){2-3} \cmidrule(lr){4-5} \cmidrule(lr){6-7} \cmidrule(lr){8-9}
   & \multicolumn{1}{c}{\textbf{Enroll}} & \multicolumn{1}{c}{\textbf{Verify}} & \textbf{EER(\%)} & \textbf{MinDCF} & \textbf{EER(\%)} & \textbf{MinDCF} & \textbf{EER(\%)} & \textbf{MinDCF}\\
  \midrule
  1 & ECAPA-TDNN$^\dag$       & ECAPA-TDNN$^\dag$     & 0.99           & 0.113           & 1.18           & 0.140           & 2.28           & 0.234 \\
  2 & ECAPA-TDNNLite$^\dag$   & ECAPA-TDNNLite$^\dag$ & 3.07           & 0.296           & 3.00           & 0.318           & 5.20           & 0.436 \\
  \midrule\midrule
  % 3 & ECAPA-TDNN$^\star$      & ECAPA-TDNN$^\star$      & 0.98           & 0.097           & 1.13           & 0.134           & 2.17           & 0.223 \\
  3 & ECAPA-TDNNLite$^\star$  & ECAPA-TDNNLite$^\star$  & 3.00           & 0.292           & 2.96           & 0.311           & 5.15           & 0.426 \\
  4 & ECAPA-TDNN$^\star$      & ECAPA-TDNNLite$^\star$  & \textbf{2.31}  & \textbf{0.251}  & \textbf{2.24}  & \textbf{0.245}  & \textbf{3.77}  & \textbf{0.358} \\
  \bottomrule
  \end{tabular}
  \label{tab:asymmetric}
\end{table*}

\section{Experimental Setup}
\label{sec:setup}
\subsection{Dataset}
Experiments are carried out on the Voxceleb dataset~\cite{nagrani2017voxceleb,chung18b_interspeech}. The development part of Voxceleb2 is employed for training, which contains 1092009 utterances from 5994 speakers. We perform online data augmentation over the training utterances with MUSAN~\cite{snyder2015musan} and RIR~\cite{7953152} datasets. There are six types of augmentation: music, babble, ambient noise, television, tempo and reverberation. The babble noise includes 3 to 8 speech files, and the television noise is a mixture of one speech file and one music file. The tempo augmentation speeds up or down utterances by 1.1 or 0.9 without changing speakers' pitch. For reverberation, we only take the small and medium simulated room impulse responses.

\subsection{Evaluation Protocol}
All systems are evaluated on clean trials of the Voxceleb1 dataset, including Voxceleb1-O, Voxceleb1-E and Voxceleb1-H. Cosine similarity is calculated between embedding pairs. Evaluation metrics include equal error rate (EER) and minimum normalized detection cost (MinDCF). $P_\text{target}$ is set to 0.01 and $C_\text{FA} = C_\text{Miss} = 1$ for MinDCF.

\subsection{Training Details}
Inputs are 80-dimensional MFCC features with 25 ms length and 10 ms shift. MFCCs are mean-normalized and no voice activity detection is applied. SpecAugment randomly masks 0 to 5 frames in both time and frequency domains of the log mel spectrograms. Last, the features are cropped into 2-second segments and 256 segments form a mini-batch. Note that segments in the same mini-batch must come from different speakers, to satisfy requirements of the AP loss.

We set the number of channels $C=512$ for ECAPA-TDNN, and $C=144$ for ECAPA-TDNNLite. The bottleneck dimension of the SE-Block and the ASP layer is 128, and the scale dimension in the SE-Res2Block equals 8. Size of the output speaker embedding is 192. AAM Softmax loss functions use a margin of 0.2 and a scale of 32. Hyperparameter $w$ in AP loss is also set to 32, and the scale factor $\lambda$ equals 10.

In the training process, model parameters update through the SGD optimizer. According to the warmup strategy, the learning rate is initialized as 0, and linearly increases to 0.1 in 5 epochs. Then the learning rate halves whenever the validation loss does not improve for over 3 epochs. The training process terminates after 100 epochs.

\section{Results}
\subsection{Performance of ECAPA-TDNNLite}
Table~\ref{tab:ecapa_tdnnlite} lists recent works on lightweight ASV models. Most of them achieve EERs in the range of 2\% and 3\%, with millions of parameters and up to 1G FLOPS. We argue that their computational costs are still expensive and thus mainly take Juliens' work as the baseline, which reaches a 3.31\% EER with only 238K parameters and 11.5M FLOPS. To have a fair comparison, we reproduce the work with the same experimental setup in Section~\ref{sec:setup}. According to experimental results, the individual ECAPA-TDNNLite model achieves a better EER of 3.07\% on Voxceleb1-O. Although our model has 30\% more parameters than Julien's, the FLOPS are the same and our model even runs 4 times faster. It shows an interesting phenomenon that the inference speed is not strictly proportional to the number of parameters or FLOPS. In our case, it is a possible reason that Julien's model stacks deep despite fewer parameters, and thus reduces the degree of parallelism. More discussions can be viewed in~\cite{MaZZS18}. We skip this topic due to space limitation.

\subsection{Performance of the Asymmetric Structure}
Performance of the asymmetric structure is shown in Table~\ref{tab:asymmetric}. Experiments 1 and 2 report EERs and MinDCFs of individual ECAPA-TDNN and ECAPA-TDNNLite models. For the rest experiments, we jointly train the two models in the asymmetric structure. Experiment 3 corresponds to the knowledge distillation method. After joint training, only the small-scale model, or namely the student model, is applied for both enrollment and verification. The performance improves by 2\% relatively in comparison with experiment 2. In the last experiment where we employ ECAPA-TDNN for enrollment and ECAPA-TDNNLite for verification, the EER metric reduces by 25\%, as well as 15\% to 23\% for MinDCF, which proves effective in comparison with the symmetric structure. The increased computational complexity only lies in the enrollment stage, which we argue is of low usage rate and less latency-sensitive to users in daily life. The asymmetric structure also shows a new solution to improve performance on IoT devices. In addition to increasing the computing capability or updating models with limited resources, it is also effective to employ a powerful model on the server for enrollment.

\begin{figure}[t]
  \centering
  \includegraphics[width=0.91\linewidth]{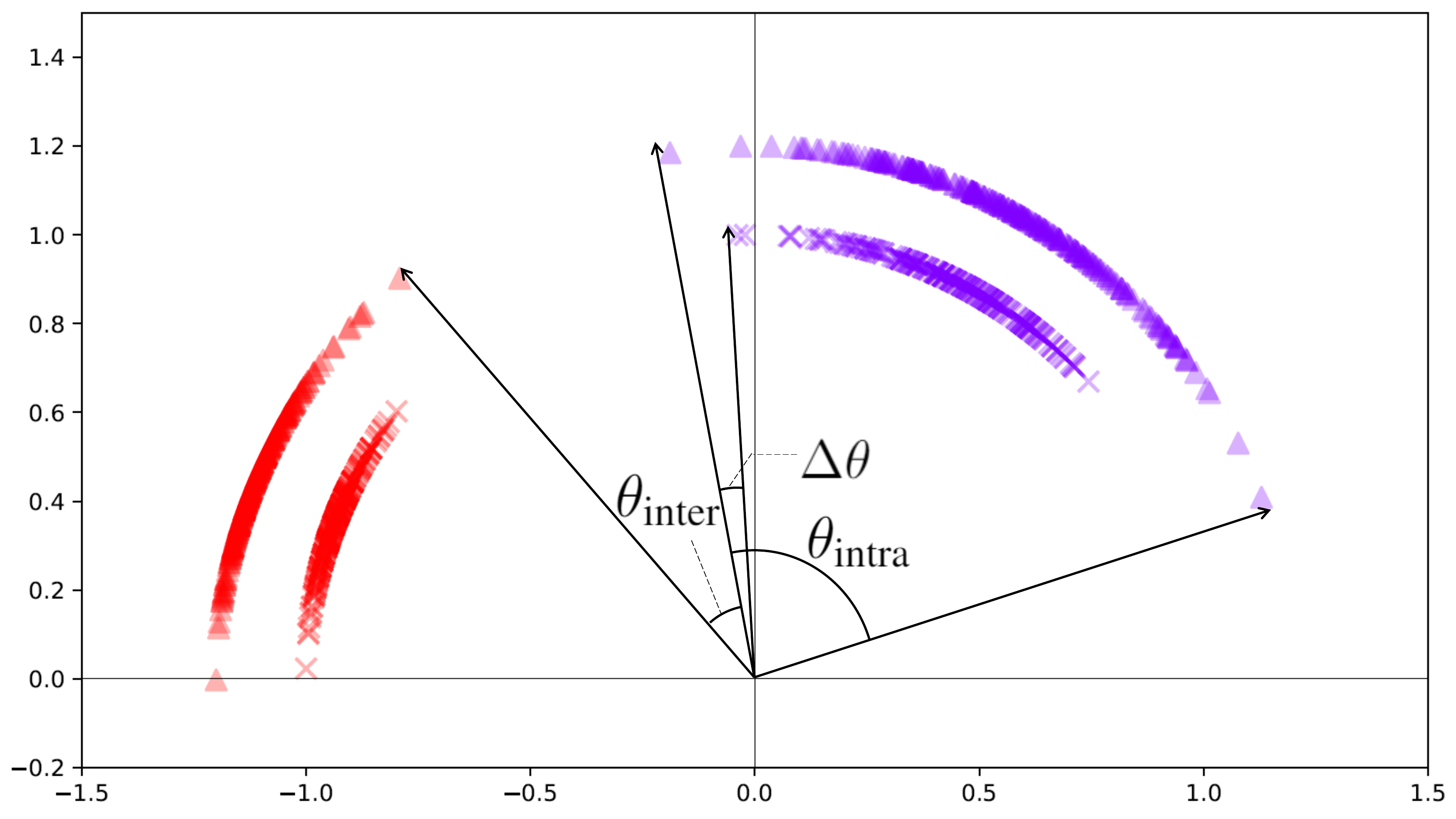}
  \caption{Explanation for effectiveness of the asymmetric structure. $\times$ and $\bigtriangleup$ denote speaker embeddings extracted by ECAPA-TDNN and ECAPA-TDNNLite, respectively. The two speakers are plotted in different colors.}
  \vspace{-3mm}
  \label{fig:analysis}
\end{figure}

To explore why the asymmetric structure is effective, we select two speakers from the Voxceleb1 test set and extract speaker embeddings with both ECAPA-TDNN and ECAPA-TDNNLite models. Then the embeddings are projected into 2D surface and normalized by different scales, as shown in Fig.~\ref{fig:analysis}. $\theta_\text{intra}$ is the maximum intra-class distance of the ECAPA-TDNNLite embeddings, and $\theta_\text{inter}$ is the minimum inter-class distance. An ASV system is expected to have smaller $\theta_\text{intra}$ and larger $\theta_\text{inter}$. When we replace the ECAPA-TDNNLite embeddings with ECAPA-TDNN embeddings for enrollment, $\theta_\text{intra}$ narrows by $\Delta\theta$ while $\theta_\text{inter}$ widens by the same value, which eventually improves the performance.

\section{Conclusions}
This paper proposes the asymmetric structure, where models of different scales are employed for enrollment and verification. ECAPA-TDNNLite is presented as the small-scale model, which achieves an EER of 3.07\% with only 11.6M FLOPS. The asymmetric structure further improves the performance by 25\% relatively, with no additional computational costs during verification.

\clearpage
\bibliographystyle{IEEEbib}
\bibliography{Template}

\end{document}